\documentclass[prb,twocolumn]{revtex4}
\usepackage{graphicx}
\usepackage{amsmath}
\usepackage[english]{babel}
\usepackage{color}

\begin{document}
\title{Beating maps of singlet fission: Full-quantum simulation of coherent two-dimensional electronic spectroscopy in organic aggregates}

\author{Ke-Wei Sun$^{1}$, Yao Yao$^{2}$\footnote{Electronic address:~\url{yaoyao2016@scut.edu.cn}}}

\affiliation{\it $^1$School of Science, Hangzhou Dianzi University, Hangzhou 310018, China\\
$^2$Department of Physics, South China University of Technology, Guangzhou 510640, China
}
\date{\today}

\begin{abstract}
The coherent two-dimensional (2D) electronic spectra with respect to the singlet fission (SF) process in organic molecular aggregates are simulated by the Davydov ansatz combined with the Frenkel-Dirac time-dependent variational algorithm. By virtue of the full-quantum dynamical approach, we are able to identify the signals of triplet excitation in the excite-state absorption contribution of the 2D spectra. In order to discuss whether a mediative charge-transfer (CT) state is necessary to SF, we increase the CT-state energy and find, in a theoretical manner, the beating signal related to the triplet is inhibited. The vibronic coherence is then studied in the beating maps for both the ground and excited state. Except for the normal beating modes adhering to the relevant electronic state, we observe signals that are explicitly related to the triplet excitations. The pathways of transition corresponding to these signals are clarified in the respective Feynman diagram, which can help the experimenters determine the physical origin of relevant measurements.
\end{abstract}

\maketitle

\section{Introduction}

Understanding of the excitonic dynamics in organic molecular aggregates has got substantial improvements in the last decade along with the massive application of the coherent electronic spectroscopy technique \cite{Chan,Plenio,Fujihashi,Monahan,Song,Butkus,Huynh,Sun}. Instead of the traditional incoherent hopping mechanism, which suits for the slow processes but can not be adapted for the ultrafast ones, several spin-free models with respect to the quantum coherence were invoked based on the delocalization \cite{Delo1}, the entropy argument \cite{Entropy} and the long-range charge-transfer (CT) state \cite{LRCT1,LRCT2,LRCT3}. In contrast, few theories on the transition between the singlet and triplet excitons, which is the major process in the singlet fission (SF), do not satisfy the vigorous growth of experiments \cite{SF0,SF1,SF2,SF3,SF4,SF5,SF6,Bakulin,SFT1,SFT2,SFT3,SFT4,SFT5,SFT6,SFT7,SFT8,SFT9,SFT10,SFT11,SFT12,SFT13,Yao,Wakasa,Th2,Th3,Th4,Zhao1,Th5,Th6,SFT14,SFT15,SFT16,Th7}. This is ascribed to the complexity of the theoretical treatment of many-body systems, that is, the singlet excitonic state, the triplet excitonic state and the CT state closely interact with each other making the quantum dynamics complicated.

The kinetic model is firstly employed to study the SF process giving the electronic structure and vibronic couplings in the crystalline tetracene and pentacene computed by the quantum chemistry methods \cite{SFT3,Th4}. By the non-adiabatic dynamical method, the basic lineshape of exciton population evolution is obtained for the two-molecule \cite{Th2,Th3} and three-molecule model \cite{SFT11}. These studies focus on the aggregates with few molecules so that the generated triplets are always correlated and bound. Wakasa \textit{et al.} tried in a different way to investigate the kinetics of the triplet pair which gives rise to a novel magnetic field effect, but in their work the mechanism of decoherence in the hopping process is not clarified \cite{Wakasa}. Afterward, more theoretical methods participate in the discussion, such as the multilayer multiconfigurational time-dependent Hartree method \cite{SFT13} and the time-dependent density matrix renormalization group (tDMRG) \cite{Yao}. One of the authors has adopted the tDMRG method to address that the SF turns out to be a completely coherent process and remarkably highlight the irreducible role of the vibronic coupling \cite{Yao}. The vibronic coherence could be experimentally visualized in the beating maps of two-dimensional (2D) electronic spectra and has been demonstrated to be essential for the excitations in organic molecules \cite{vc1,vc2}. On the basis of this critical point, it is thus a motivated subject to study the coherent 2D spectrum of SF process in organic aggregates.

The experimental technique of coherent 2D electronic spectroscopy serves as a powerful tool for measuring the quantum coherence in the molecular materials \cite{Cho11}. It is quite straightforward to apply this technique to the coherent dynamics of SF. A recent experiment has uncovered the beating maps associated with the coherent transition between the ground state and the multiexcitonic state which obviously reflect the behavior of vibronic coherence \cite{Bakulin}. Although the beating maps have been simulated on the Redfield level by the authors of the paper, the fine details of the experiment have not been explicitly rebuilt which are significant for us to properly understand the physics inside. Another computation of the 2D spectrum done by Tempelaar and Reichman focused on the correlated triplet pairs \cite{Th5}. In that work, however, neither the signals of the stimulated emission (SE) nor the vibronic coherence are present, and due to the limitation of the method, the evolution time is not sufficiently long to cover all the frequencies of interest. In this context, an efficient full-quantum simulating approach with minority assumptions is thus demanded, and subsequently, the motivation of the present work is to adopt the Davydov ansatz method, which is fast and efficient in the computational manner, to study the 2D electronic spectra and beating maps related to the SF process. The rest of the paper is organized as follows: Section II shows the model and the methodology employed in this work. In Section III the simulation results of 2D spectra and the beating maps are given, and relevant discussions are addressed. A brief summary is present in last Section.

\section{METHODOLOGY}

As a normal consideration in studying the 2D electronic spectrum, a model consisting of both the system and the harmonic bath is employed, whose Hamiltonian could be written as \cite{Bakulin,Sun,Yao}
\begin{equation}\label{total hamiltonian}
H=H_{\rm S}+H_{\rm B}+H_{\rm S-B}.
\end{equation}
Herein, the first term is the Hamiltonian of the system for studying the SF which takes excitonic states under investigation and linear vibronic couplings to relevant phonon modes. The form of $H_{\rm S}$ reads \cite{Huynh,Sun},
\begin{equation}\label{hamiltonian1}
H_{\rm S}=H_{\rm ex}+H_{\rm ph}+H_{\rm ex-ph},
\end{equation}
where $H_{\rm ex}$ represents a widely-studied Frenkel exciton (FE)-charge transfer (CT) mixing Hamiltonian\cite{Yao}, with the form being
\begin{eqnarray}\label{hamiltonian2}
H_{\rm ex}=\sum_{i}\epsilon_i|i\rangle\langle i|+(J_1|{\rm S}_1\rangle\langle {\rm CT}|+J_2|{\rm TT}_1\rangle\langle {\rm CT}|+{\rm h.c.}),
\end{eqnarray}
where $i$ labels the electronic states in order, namely $i\in\{{\rm S_0,S_1,CT,TT_1,S}_n,{\rm TT}_n\}$; $\epsilon_i$ denotes the energy of the respective state with the ground-state energy setting to zero; $J_1$ and $J_2$ represent the effective couplings for the charge transferring from S$_1$ and TT$_1$ to the mediative CT state, respectively. Herein, we have three manifolds of states: the ground-state manifold `g' including the state ${\rm S}_0$, the first excited manifold `e' including ${\rm S_1,CT}$ and ${\rm TT_1}$, and the higher-lying excited manifold `f' including ${\rm S}_n$ and ${\rm TT}_n$. The higher-lying excited states are taken into account in order to study the excite-state absorption (ESA) in the spectra. So far, the value of the index $n$ can not be explicitly determined in terms of the complicated higher-lying excited manifold of organic molecules, so that we set the energy of these states with respect to the relevant experimental measurement \cite{Bakulin,Th5}. In the Hamiltonian (\ref{hamiltonian1}), $H_{\rm ph}$ represents the primary vibrational modes coupling to the excitons and the form is given by
\begin{equation}
H_{\rm ph}=\sum_q\hbar\omega_qb^\dag_qb^{}_q,
\end{equation}
where $b_q$ ($b_q^\dag$) denotes the annihilation (creation) operator of the vibrational mode with frequency $\omega_q$. $H_{\rm ex-ph}$ is for the electron-phonon (vibronic) coupling with the diagonal form\cite{Bakulin}
\begin{equation}
H_{\rm ex-ph}=\sum_{q}\hbar\omega_q(b_q^{\dag}+b_q)\sum_{i}\frac{\Delta^i_q}{\sqrt{2}}|i\rangle\langle i|,
\end{equation}
where $\Delta^{i}_q$ denotes the respective coupling strength. Throughout this work, we take $\Delta^{S_0}_q=0$ and all other $\Delta^{i}_q$ are taken to be the same as $\Delta_q$ .

A secondary harmonic phonon bath is involved in the Hamiltonian~(\ref{total hamiltonian}) for generating the dephasing in the spectrum \cite{Huynh,Sun}, in which the second and the third terms are those for the bath and the system-bath interaction, respectively, that is
\begin{equation}\label{s-b1}
H_{\rm B}=\sum_{\mu}\hbar\Omega_{\mu}B^\dag_{\mu}B^{}_{\mu},
\end{equation}
and
\begin{equation}\label{s-b2}
H_{\rm S-B}=\sum_{\mu}\hbar\Omega_{\mu}\left(B^{\dag}_{\mu}+B_{\mu}\right)\sum_{i}\kappa_{\mu}|i\rangle\langle i|,
\end{equation}
where $B^{}_{\mu}$ ($B^\dag_{\mu}$) is the annihilation (creation) operator of the secondary phonon in the bath with frequency $\Omega_{\mu}$, and $\kappa_{\mu}$ is the respective coupling strength. The bath spectral density is given as $D(\omega)=\sum_j\kappa_j^2\Omega_j^2\delta(\omega-\Omega_j)$. For simplicity, we have assumed that the system-bath coupling is all the same for excited states, so that the bath degrees of freedom can be traced out analytically yielding an exact master equation for the reduced (system) density matrix \cite{Huynh,Sun}.

In order to simulate the 2D photo echo (PE) spectrum, we have to additionally consider the interaction between the system and the light field. The corresponding Hamiltonian is given by $H_{\rm L}=-({\mathbf E}({\mathbf r},t)\cdot\hat{\bf{\mu}}_++{\mathbf E}^*({\mathbf r},t)\cdot\hat{\bf{\mu}}_-)$, with ${\mathbf E}({\mathbf r},t)$ being the time-dependent electric field of the applied pulse sequence. Herein, $\hat{\mu}_+$ denotes the excitation operator which is defined as
\begin{eqnarray}\label{s-b3}
\hat{\mu}_+=\mu\left(|{\rm S}_1\rangle\langle {\rm S}_0|+|{\rm S}_n\rangle\langle {\rm S}_1|+2.5|{\rm TT}_n\rangle\langle {\rm TT}_1|\right),
\end{eqnarray}
with $\mu$ being the transition dipole moment, and the detection operator $\hat{\mu}_-$ is conjugated to $\hat{\mu}_+$. Here in this work we regard the CT and TT$_1$ states to be dark states. The transition dipole moment of TT$_1$ and TT$_n$ is larger than that of singlet because of the relatively strong triplet absorption \cite{Bakulin}.

We employ the full-quantum Davydov ansatz method to calculate the dynamics and then the 2D PE spectum (see the Appendix for the theoretical details). The Davydov ansatz method has demonstrated itself to be sufficiently efficient to compute the long-term evolution taking quantum phonons into account \cite{Davydov1,Davydov2,Sunkw,Sun}, so that we are able to obtain more information on a wider frequency regime than that in the previous works \cite{Bakulin,Th5}. The parameters are set as follows \cite{Bakulin}. The energies are taken as $\epsilon_{\rm S_1}=15560\rm{cm^{-1}}$,
$\epsilon_{\rm CT}=19431.5\rm{cm^{-1}}$, $\epsilon_{\rm TT_1}=14780\rm{cm^{-1}}$,
$\epsilon_{{\rm S}_n}=31000\rm{cm^{-1}}$ and $\epsilon_{{\rm TT}_n}=29560\rm{cm^{-1}}$.
Herein, the CT-state energy $\epsilon_{\rm CT}$ is computed by considering the electron and hole are residing in the nearest molecules \cite{LRCT3}.
In the real case, however, the CT-state could be of long range and $\epsilon_{\rm CT}$ is variable, so that we will consider to adjust it in an extent as discussed below.
The transfer integral $J_1=J_2=800\rm{cm^{-1}}$ such that the eigen-energies of the first excited manifold are given by
$\epsilon_{e_1}=14625\rm{cm^{-1}}$, $\epsilon_{e_2}=15432\rm{cm^{-1}}$, and $\epsilon_{e_3}=19715\rm{cm^{-1}}$. Comparing the energies,
one can find the diabatic state TT$_1$ has got the overwhelming weight of the adiabatic state e$_1$ while the S$_1$ contributes mostly to the state e$_2$.
The most pronounced vibrational modes participating in the SF process are found to be $\omega_1=265\rm cm^{-1}$, $\omega_2=1170\rm cm^{-1}$
and $\omega_3=1360\rm cm^{-1}$ observed by the resonance Raman spectra in the crystalline pentacene film \cite{Bakulin}. The respective vibronic couplings
are $\Delta_1=0.47$, $\Delta_2=0.4$, $\Delta_3=0.5$.

\section{Results and Discussion}
\subsection{2D spectra}

\begin{figure}
\includegraphics[scale=0.3,trim=0 60 0 40]{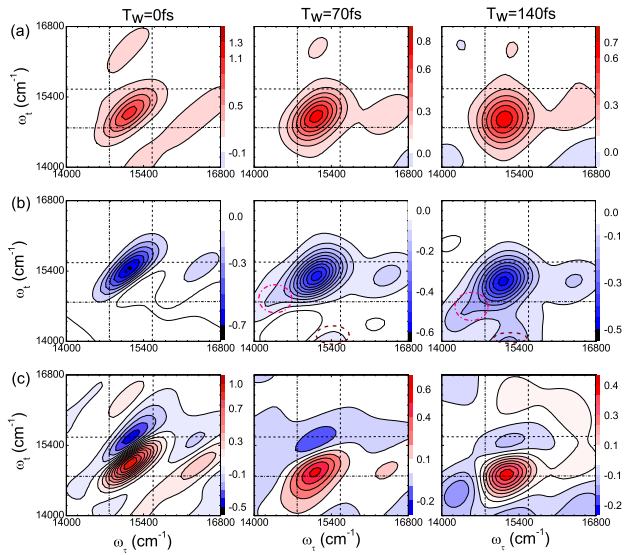}
\caption{2D PE spectrum $S(\omega_\tau,T_w,\omega_t)$ at three waiting times $T_{\rm w}=0$, $70$ and $140$fs from the first to the third column. (a) shows the GSB+SE contribution, (b) shows the ESA contribution, and (c) shows the combined one. The energies of S$_1$ and TT$_1$ state are indicated by dash and dash-dot lines, respectively. The wine dashed circles in the second and the third columns of (b) figure out the emergence of negative peak for the triplet absorption, and the pink dash-dotted circles indicate negative peak at around $\omega_{\tau}\sim\epsilon_{e_1}$, $\omega_t\sim\epsilon_{{\rm TT}_n}-\epsilon_{e_1}$.
}\label{figure1}
\end{figure}
The real part of the 2D PE spectrum $S(\omega_\tau,T_w,\omega_t)$ at the different waiting time $T_{\rm w}$ is plotted in Fig.~\ref{figure1}, where the contribution of the ground-state bleaching (GSB) plus the SE is displayed in (a), the ESA contribution is in (b), and the combined one is in (c). In Fig.~\ref{figure1}(a), it is found at $T_{\rm w}=0$ a main positive diagonal peak arises at around $(\omega_{\tau}=15139{\rm cm}^{-1},\omega_t=15139{\rm cm}^{-1})$. This peak stems from the vertical transition of singlet states. Compared with the energy of e$_2$ state, there is a red shift of about $293{\rm cm}^{-1}$ which is originated from the presence of the vibrational modes that yield a reorganization energy $\lambda_i=\frac{1}{2}\sum_{q}(\Delta^{i}_q)^2\omega_q$. The red shift will always be present in the following figures. In addition, two nearly symmetric off-diagonal peaks emerge from the absorption and the emission between the ground and first excited manifold accompanying with some pronounced vibrational structures \cite{Egorova1}. As the waiting time $T_{\rm w}$ increases, the elongation and the tilt of the peak become rounded and less pronounced because of the dephasing effect from the bath. Moreover, the main peak shifts downward due to the spectral diffusion process \cite{Cho11}.

The negative peaks shown in Fig.~\ref{figure1}(b) are of more importance which reflect the rich information of the excited-state manifolds. The two negative peaks which clearly present at the different excitation frequencies $15139{\rm cm}^{-1}$ and $16430{\rm cm}^{-1}$ show the transitions of g to e$_2$ and e$_2^{\prime}$, with the prime denoting the relevant vibronic state. These two peaks share the same probe frequency at $\omega_t\sim15455{\rm cm}^{-1}$ representing the created coherence between the electronic excited states e$_2$ and S$_n$ after the third matter-light interaction in the relevant experiment (see the Feynman diagram below).

With $T_{\rm w}$ increasing, a small negative peak indicated by the dashed circle gradually grows up to connect the main negative peak at $(15139{\rm cm}^{-1},15455{\rm cm}^{-1})$. It stems from the population transferring process $|\rm{e_2\rangle\langle e_2|\to|e_{1}^{\prime}\rangle\langle e_{1}^{\prime}|\to|e_1\rangle\langle e_1|}$ (namely the SF process) during the waiting period, and the coherence between the electronic excited states TT$_n$ and e$_1^{\prime}$(e$_1$) could be presented during the detection period. Moreover, one would observe the generation of another negative peak as indicated by the dash-dotted circle in the ESA signal. The peak manifests a nonzero (but weak) transition dipole moment of the states g and ${\rm e}_1$, and the latter has a majority TT$_1$ (dark state) component. This is because the S$_1$ and T$_1$ states indirectly couple to each other mediated by the CT state. These two peaks discussed above, both correlated with the triplet excitons, manifest clear signals that SF process does take place. These pronounced signals are successfully obtained benefitting from the advantages of dealing with the full-quantum dynamics in our simulations \cite{Th5}.

The complete spectrum at different waiting times is plotted in Fig.~\ref{figure1}(c). The positive peaks partially cancel the negative peaks, and the ESA signal from singlets is slowly lost due to the SF process \cite{Bakulin}. It is worth noting that no obvious absorption feature about the CT state is observed in the 2D spectrum, implying that the CT state has a negligible transition dipole in the model.

\subsection{Mediative CT state}

\begin{figure}
\includegraphics[scale=0.3,trim=50 100 0 100]{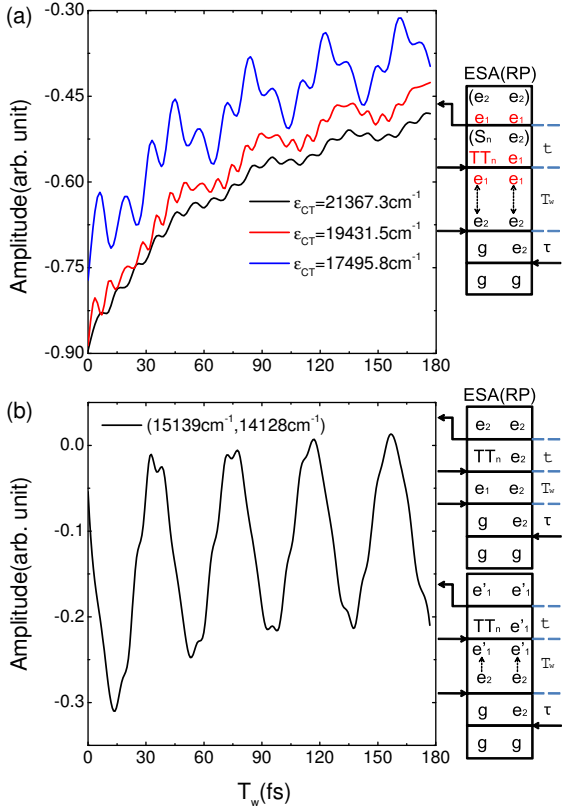}
\caption{(a) Time traces of the amplitude of the main negative peaks with three CT-state energies. (b) Time traces of the amplitude of the negative peak at (15139$\rm cm^{-1}$,14128$\rm cm^{-1}$) for $\epsilon_{\rm CT}=19431.5\rm cm^{-1}$. On the right hand side of each panel, the Feynman diagrams with respect to the relevant pathway are shown.
}\label{figure2}
\end{figure}

It is hotly debated whether we need a CT state to mediate the transition between singlet and triplet. Fig.~\ref{figure2}(a) therefore displays the temporal dependence of the amplitudes of the main negative peaks in ESA signal (absorption signal of singlet exciton) with three CT-state energies. On the experimental side, the CT-state energy might be adjusted by, for example, changing the intermolecular distance \cite{Th6}. One can find here that, all the three curves tend to decay from some negative value to zero with increasing $T_{\rm w}$ due to the dephasing effect. The higher the CT-state energy, the slower the dephasing. More importantly, a weaker oscillation is observed for the higher CT-state energy. From the corresponding Feynman diagram shown on the right hand side of the panel, the oscillation stems from population transferring between $|\rm{e}_2\rangle\langle {\rm e}_2|$ and $|{\rm e}_{1}\rangle\langle {\rm e}_{1}|$ which represents the SF process and the geminate fusion process (namely the triplet pair returns to the singlet exciton). As a result, it implies that with the CT-state energy increasing the singlet-triplet transition will be inhibited, which can serve as a useful fingerprint for experimenters observing the SF.

Fig.~\ref{figure2}(b) shows the behavior of the negative peak at $(15139{\rm cm}^{-1},14128{\rm cm}^{-1})$ for $\epsilon_{\rm CT}=19431.5\rm cm^{-1}$, where the quantum beating effect is exhibited. The peak intensity oscillates with a period of $T\sim 40$fs, from which we can get the frequency of the beating mode is $\omega\sim 834\rm cm^{-1}$. This oscillation mainly arises from the contribution of the ESA with the wave vector of the light being ${\mathbf k}_I=-{\mathbf k}_1+{\mathbf k}_2+{\mathbf k}_3$, with which the electronic coherence between e$_1$ and e$_2$ states ($\epsilon_{e_2}-\epsilon_{e_1}=807\rm cm^{-1}$) is formed during the waiting time. In addition, the other possible contribution stems from the population transfer $|\rm{e_2\rangle\langle e_2|\to|e_{1}^{\prime}\rangle\langle e_{1}^{\prime}|}$ during the waiting time. All these pathways are given in the Feynman diagrams of Fig.~\ref{figure2}(b).

\subsection{Beating maps}

One has been noticed that the vibronic state matters a lot in the SF process, so it is quite worthwhile to investigate the beating maps which serve as a powerful tool to distinguish GSB, SE, and ESA contributions to the vibrational signal component. Generally speaking, GSB signal carries the information of the ground-state coherence, and SE and ESA signals reflect the excited-state coherence. The definition of the beating map is on the basis of the Fourier transformation of the 2D PE signal over the waiting time $T_{\rm w}$, which is given by\cite{Bakulin,Egorova2}
\begin{eqnarray}\label{beating}
S_{\rm R(NR)}(\omega_\tau,\omega_{T},\omega_t)&=&{\mathrm{Re}}\int_{0}^{\infty}dT_{\rm w}\cdot iP^{(3)}_{\rm R(NR)}(\omega_\tau,T_{\rm w},\omega_t)\nonumber\\
&\times&\exp(i\omega_{T}T_{\rm w}).
\end{eqnarray}
As a normal treatment, we draw the absolute value of $S_{\rm R(NR)}$ in Eq.~(\ref{beating}) in the beating maps.

\begin{figure}
\includegraphics[scale=0.75]{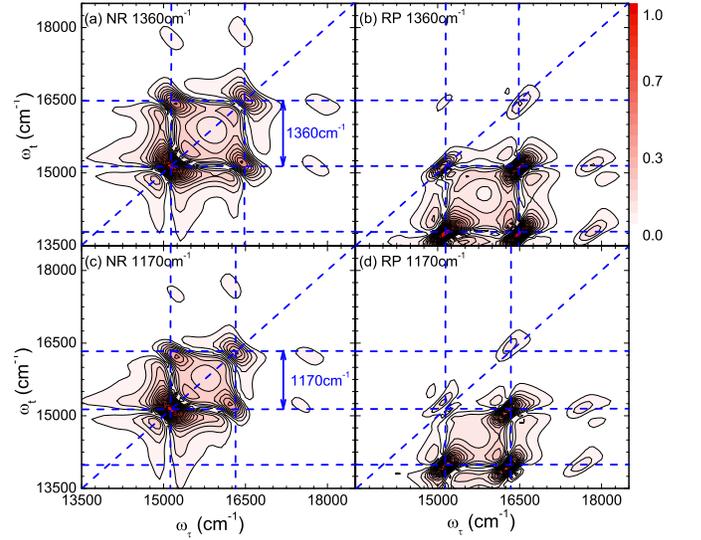}
\caption{Nonrephasing and rephasing beating maps for the ground state with $\omega_{T}$ being 1360$\rm cm^{-1}$ and 1170$\rm cm^{-1}$.
}\label{figure3}
\end{figure}

The nonrephasing and rephasing beating maps for the ground-state coherence are shown in Fig.~\ref{figure3}, with $\omega_{T}$ being the two primary
frequencies of the vibrational modes, i.e., 1360$\rm cm^{-1}$ and 1170$\rm cm^{-1}$. The frequency we calculate covers a wider regime of interest than that in the previous work such that more pronounced structures are discovered \cite{Th5}. One can find that, the nonrephasing maps are diagonally symmetric with the frequency difference of the peaks being $\omega_{T}$, since the vibrations are in the vacuum state on both the initial and final stages of the matter-light interacting process. On the other hand, the rephasing maps are obtained by shifting the corresponding nonrephasing ones to the lower probe frequencies $\omega_t$ by the corresponding beating mode frequency $\omega_{T}$, implying on the final stage the vibrations do not relax to their vacuum state along with the electronic ground state. By analyzing the pathway of each map, it is summarized as follows. In the nonrephasing maps, the vibronic coherence with respect to the ground state gives rise to the four peaks of g to e$_2$ and g to e$_2^{\prime}$ excitation and e$_2$ to g and e$_2^{\prime}$ to g detection (closed loops), while the four peaks of g to e$_2$ and g to e$_2^{\prime}$ excitation and e$_2$ to g$^{\prime}$ and e$_2^{\prime}$ to g$^{\prime}$ detection are present in the rephasing maps. Due to the relatively weak transition dipole moment ($\mu_{e_1g}\sim0.17\mu_{e_2g}$), we do not find the signals with e$_1$(e$_1^{\prime}$) excitation in the GSB contributions, implying the TT signal can not be observed in the GSB beating maps.

\begin{figure}
\includegraphics[scale=0.75]{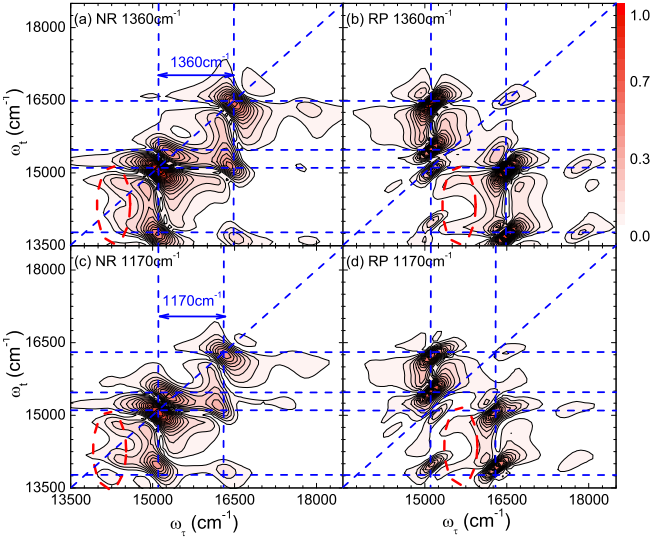}
\caption{Nonrephasing (NP) and rephasing (RP) beating maps for the excited states with $\omega_{T}$ being 1360$\rm cm^{-1}$ and 1170$\rm cm^{-1}$. The red dashed circles indicate the g to e$_1$ excitation and TT$_n$ (TT$_n^{\prime}$) to e$_1^{\prime}$ detection.
}\label{figure4}
\end{figure}

\begin{figure}
\includegraphics[scale=0.33,trim=30 50 0 50]{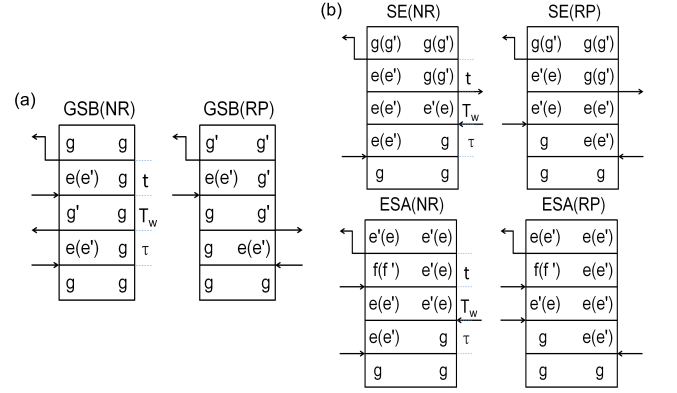}
\caption{Nonrephasing and rephasing Feynman diagrams for the beating maps of (a) the ground-state coherence and (b) the excited-state coherence.
}\label{figure5}
\end{figure}

The nonrephasing and rephasing beating maps for the excited-state coherence formed by SE and ESA contributions are shown in Fig.~\ref{figure4}, which provide more profound structures. Similar with that in Fig.~\ref{figure3}, one can find several main peaks resulting from the e$_2$(e$_2^{\prime}$) excitations, which can approximately transform into each other upon the exchange of the two excitation frequencies \cite{Egorova2}. The interval of the excitation frequencies of the main peaks $\omega_{\tau}$ is exactly the beating-mode frequency $\omega_{\rm T}=$1360$\rm cm^{-1}$ or 1170$\rm cm^{-1}$. In order to facilitate the analysis of the pathways of the intrinsic process, we draw the corresponding Feynman diagrams for nonrephasing (NR) and rephasing (RP) spectral in Fig.~\ref{figure5}. In the third row of each Feynman diagram, it shows the state during the waiting time $T_{\rm w}$ after the first two actions of light-matter interactions, and one can find the term of coherence between g and g$'$ states (ground-state coherence) in the GSB diagram and e and e$'$ states (excited-state coherence) in the SE and ESA diagrams. The GSB contributions for the beating maps have been discussed above. For SE contributions, the corresponding main peaks arise from g to e$_2$(e$_2^{\prime}$) excitation and e$_2$(e$_2^{\prime}$) to g or g$^{\prime}$ detection in the nonrephasing maps, while in the rephasing maps the main peaks step from g to e$_2$(e$_2^{\prime}$) excitation and e$_2^{\prime}$(e$_2$) to g or g$^{\prime}$ detection. For ESA contributions, the corresponding main peaks arise from g to e$_2$(e$_2^{\prime}$) excitation and f or f$^{\prime}$ to e$_2^{\prime}$(e$_2$) detection in the nonrephasing maps, while in the rephasing maps the main peaks step from g to e$_2$(e$_2^{\prime}$) excitation and f or f$^{\prime}$ to e$_2$(e$_2^{\prime}$) detection. These processes corresponds to the main peaks of beating maps in Fig.~\ref{figure4}.

As we are mainly concerning the SF process, the signals that are related to the triplet excitons are of significance. Remarkably, in the ESA contributions there are some weak but observable signals with e$_1$(e$_1^{\prime}$) excitations as indicated by the dashed circles in  Fig.~\ref{figure4}(a) and (c). As discussed in the 2D spectra above, these TT-relating signals emerge because the transition dipole moment of TT is larger than that of S$_1$, namely $\mu_{TT_1TT_n}=2.5\mu_{S_1g}$, which gives rise to the processes of g to e$_1$ excitation and TT$_n$ or TT$_n^{\prime}$ to e$_1^{\prime}$ detection \cite{note1} in the NR diagrams. In addition, the peaks of e$_1^{\prime}$(e$_1$) excitation are observed as well in the RP maps, as indicated by the dashed circles in Fig.~\ref{figure4}(b) and (d). Our results thus suggest that the vibronic coherence in the beating maps from ESA contributions can be measured as a fingerprint with respect to the triplet pair in the SF process.

\section{Conclusion}

In summary, we have adopted a full-quantum approach to simulate the coherent 2D spectra considering a benchmarking FE-CT mixing model for the SF process. The vibrational degrees of freedom are also introduced in the model since the vibronic coherence usually plays an important role in the ultrafast SF process. The pronounced vibrational structures in the spectra and the quantum beating pictures are observed in our calculations, and the SF process could be identified by virtue of studying the evolution of the amplitudes of the relevant peaks. In addition, the ground- and excited-state vibronic coherences arising from the vibrations coupling to electronic states are separately investigated in beating maps. The TT-relating signals in the beating map with ESA contribution essentially act as a fingerprint for the SF process.

\section*{Acknowledgment}
The authors gratefully acknowledge support from the National Natural Science Foundation of China (Grant Nos.~11404084, 91333202 and 11574052). We thank Y. Fujihashi for helpful discussions.

\appendix
\section{Wavefunction formulation}
We adopt the Davydov $\rm {D}_1$ ansatz to study the dynamics of the SF process. The initial state of the system is assumed to be the electronic ground state $|S_0\rangle$ ($|g\rangle$) with the relevant vibrational ground state $|0\rangle_{\rm ph}$, i.e., $|\Psi_{0}\rangle=|S_0\rangle|0\rangle_{\rm ph}$. The time-dependent wavefunction of the excited-state manifold reads\cite{Davydov1,Davydov2,Sunkw,Sun}
\begin{equation}\label{localDavyAnsatz}
\big|\Psi_{\rm D_1}^{e(f)}(t)\rangle = \sum_{i\in{|e(f)\rangle}}A_{i}(t)|i\rangle|\lambda_{i,q}(t)\rangle_{\rm{ph}},
\end{equation}
and
\begin{equation}
|\lambda_{i,q}(t)\rangle_{\rm{ph}}=\exp\Big\{\sum_{q}\big[\lambda_{i,q}(t)\hat{b}_{q}^{\dagger}-{\rm H.c.}\big]\Big\}|0\rangle_{\rm ph},
\end{equation}
where $A_i(t)$ and $\lambda_{i,q}(t)$ are the variational parameters which could be derived from the Dirac-Frenkel time-dependent variational principle as expressed below.

The details of the variational procedure can be described as follows. We first write down the variational equations as\cite{Sunkw,Sun,Huynh}
\begin{equation}
\frac{d}{dt}(\frac{\partial L^{e(f)}}{\partial\dot{A}_{i}^{*}})-\frac{\partial L^{e(f)}}{\partial A_{i}^{*}}=0 \label {Lagrangian1},
\end{equation}
\begin{equation}
\frac{d}{dt}(\frac{\partial L^{e(f)}}{\partial\dot{\lambda}_{i,q}^{*}})-\frac{\partial L^{e(f)}}{\partial\lambda_{i,q}^{*}}=0, \label {Lagrangian3}
\end{equation}
where the Lagrangian $L^{e(f)}$ of the system is formulated as
\begin{eqnarray}\label {Lagrangian4}
L^{e(f)}&=&\langle
 {\Psi_{\rm D_1}^{e(f)}}(t)|{\frac{i\hbar}{2}}\frac{\overset{\longleftrightarrow}{\partial}}{\partial
t}-\hat{H}|{\Psi_{\rm D_1}^{e(f)}}(t)\rangle  \nonumber \\
&=&\frac{i\hbar}{2}[\langle
{\Psi_{\rm D_1}^{e(f)}}(t)|\frac{\overrightarrow{\partial}}{\partial
t}|{\Psi_{\rm D_1}^{e(f)}}(t)\rangle-\langle
{\Psi_{\rm D_1}^{e(f)}}(t)|\frac{\overleftarrow{\partial}}{\partial
t}|{\Psi_{\rm D_1}^{e(f)}}(t)\rangle] \nonumber \\
&&-\langle{\Psi_{\rm D_1}^{e(f)}}(t)|\hat{H}|{\Psi_{\rm D_1}^{e(f)}}(t)\rangle.
\end{eqnarray}
Based on Eqs.~(\ref{Lagrangian1})-(\ref{Lagrangian4}), the time-dependent wavefunction $|\Psi^{e(f)}_{\rm D_1}(t)\rangle$ is derived. As a result, the equations of motion for the time-dependent variational parameters $A_i(t)$ and $\lambda_{i,q}(t)$ in the excited-state manifold e are expressed as
\begin{eqnarray}
-i\dot{A}_i&=&\frac{i}{2}A_i\sum_q(\dot{\lambda}_{i,q}\lambda_{i,q}^*-c.c.)-\sum_{j\neq i}J_{ij}A_jS_{i,j}\nonumber\\
&&-\epsilon_iA_i+\sum_q\frac{\omega_q}{\sqrt{2}}\Delta^i_qA_i(\lambda_{i,q}+c.c.)\nonumber\\
&&-\sum_q\omega_qA_i|\lambda_{i,q}|^2,
\end{eqnarray}
\begin{eqnarray}
iA_i\dot{\lambda}_{i,q}&=&\sum_{j\neq i}J_{ij}A_j(\lambda_{j,q}-\lambda_{i,q})S_{i,j}\nonumber\\
&&-\Delta_q^i\frac{\omega_q}{\sqrt{2}}A_i+\omega_q\lambda_{i,q}A_i,
\end{eqnarray}
with the Debye-Waller factor being
\begin{eqnarray}
S_{i,j}=\exp[-\frac{1}{2}\sum_{q}(|\lambda_{i,q}|^2+|\lambda_{j,q}|^2-2\lambda_{i,q}^*\lambda_{j,q})],\nonumber\\
\end{eqnarray}
where $J_{{\rm S}_1,{\rm CT}}=J_{{\rm CT},{\rm S}_1}=J_1$ and $J_{\rm CT,TT}=J_{\rm TT,CT}=J_2$, and the other coupling parameters $J_{ij}$ are set to zero in our model. For the higher excited-state manifold f, the equations of motions are similarly written as
\begin{eqnarray}
-i\dot{A}_i&=&\frac{i}{2}A_i\sum_q(\dot{\lambda}_{i,q}\lambda_{i,q}^*-{\rm c.c.})-\epsilon_iA_i-\sum_q\omega_qA_i|\lambda_{i,q}|^2\nonumber\\
&+&\sum_q\frac{\omega_q}{\sqrt{2}}\Delta^i_qA_i(\lambda_{i,q}+{\rm c.c.}),
\end{eqnarray}
\begin{eqnarray}
iA_i\dot{\lambda}_{i,q}=-\Delta_q^i\frac{\omega_q}{\sqrt{2}}A_i+\omega_q\lambda_{i,q}A_i.
\end{eqnarray}

\section{Third-order response functions}
As described in the main text, the light-matter interaction Hamiltonian is given by
\begin{equation}\label{H_L}
H_{\rm L}=-\left({\mathbf E}({\mathbf r},t)\cdot\hat{\bf{\mu}}_+
+{\mathbf E}^*({\mathbf r},t)\cdot\hat{\bf{\mu}}_-\right),
\end{equation}
where ${\mathbf E}$ is the external electric field and can be described as
\begin{eqnarray}
\textbf{E}(\textbf{r},t)&=&\textbf{E}_1(\textbf{r},t)+\textbf{E}_2(\textbf{r},t)+\textbf{E}_3(\textbf{r},t)\nonumber\\
\textbf{E}_1(\textbf{r},t)&=&\textbf{e}_1E_1(t-\tau_1)e^{i\textbf{k}_1\cdot\textbf{r}-i\omega_1t+i\phi_1}\nonumber\\
\textbf{E}_2(\textbf{r},t)&=&\textbf{e}_2E_2(t-\tau_2)e^{i\textbf{k}_2\cdot\textbf{r}-i\omega_2t+i\phi_2}\nonumber\\
\textbf{E}_3(\textbf{r},t)&=&\textbf{e}_3E_3(t-\tau_3)e^{i\textbf{k}_3\cdot\textbf{r}-i\omega_3t+i\phi_3},
\end{eqnarray}
where $\textbf{e}_a$, $\textbf{k}_a$, $\omega_a$, $E_a(t)$, and $\phi_a$ ($a=1,2,3$) denote the polarization, the wave vector, the frequency, the dimensionless envelope, and the initial phase. We then define the pulse arrival time in the system-field Hamiltonian (\ref{H_L}) as
\begin{equation}
\tau_{1}=-T_{\rm w}-\tau,\,\,\,\tau_{2}=-T_{\rm w},\,\,\,\tau_{3}=0,\label{Ttau}
\end{equation}
where $\tau$ (the so-called coherence time) is the delay time between the second and the first pulse, and $T_{\rm w}$ (the so-called population time) is the delay time between the third and the second pulse. In the short pulse limit, we have $E_a(t)=E_0\delta(t)$.

In order to evaluate the 2D PE spectra, four contributions to the third-order response function have to be calculated, which can be expressed in terms of four-time correlation functions as follows \cite{Abramavicius,Mukamel,Sun}:
\begin{equation}\label{R1}
R_{1}(t_{3},t_{2},t_{1})=\Phi(t_{1},t_{1}+t_{2},t_{1}+t_{2}+t_{3},0),
\end{equation}
\begin{equation}\label{R2}
R_{2}(t_{3},t_{2},t_{1})=\Phi(0,t_{1}+t_{2},t_{1}+t_{2}+t_{3},t_{1}),
\end{equation}
\begin{equation}\label{R3}
R_{3}(t_{3},t_{2},t_{1})=\Phi(0,t_{1},t_{1}+t_{2}+t_{3},t_{1}+t_{2}),
\end{equation}
\begin{equation}\label{R4}
R_{4}(t_{3},t_{2},t_{1})=\Phi(t_{1}+t_{2}+t_{3},t_{1}+t_{2},t_{1},0),
\end{equation}
where the auxiliary correlation function $\Phi$ equals to the sum of two contributions $\Phi^{\rm{e}}$ (from the excited manifold e) and $\Phi^{\rm{f}}$ (from the higher-excited manifold f). These two contributions have the form as
\begin{eqnarray}\label{ms}
\Phi^{e}(\tau_{4},\tau_{3},\tau_{2},\tau_{1})=
\left\langle
\hat{\mu}_-(\tau_{4})\hat{\mu}_+(\tau_{3})\hat{\mu}_-(\tau_{2})\hat{\mu}_+(\tau_{1})
\right\rangle
\end{eqnarray}
and
\begin{eqnarray}\label{md}
\Phi^{f}(\tau_{4},\tau_{3},\tau_{2},\tau_{1})=
\left\langle
\hat{\mu}_-(\tau_{4})\hat{\mu}_-(\tau_{3})\hat{\mu}_+(\tau_{2})\hat{\mu}_+(\tau_{1})
\right\rangle,
\end{eqnarray}
where $\hat{\mu}_{\pm}(\tau)$ has got the Heisenberg representation of $\hat{\mu}_{\pm}$. $\left\langle ...\right\rangle \equiv  {\rm Tr}_{\rm B} \{\rho_{\rm B}
\langle\Psi_{0}| ...
|\Psi_{0}\rangle \}$, where $\rho_{\rm B}=Z^{-1}_{\rm B}\exp\{-\beta H_{\rm B}\}$
is the equilibrium distribution over the bath phonons at the temperature $T_{\rm eq}$. Herein, $Z_{\rm B}$ is the partition function, $\beta=(k_{\rm B} T_{\rm eq})^{-1}$, and $k_{\rm B}$ is the Boltzmann constant.

The dynamics governed by $H_{\rm S}$ and $H_{\rm SB}+H_{\rm B}$ is separable because $H_{\rm S}$ commutes with the bath Hamiltonian $H_{\rm B}$ and the system-bath coupling $H_{\rm SB}$, implying the bath is merely responsible for electronic dephasing. The correlation functions can thus be formed as a product of the system and bath counterparts \cite{Huynh,Sun}. Subsequently, we obtain
\begin{eqnarray}\label{rf}
\Phi^{e(f)}(\tau_{4},\tau_{3},\tau_{2},\tau_{1})=F^{e(f)} (\tau_{4},\tau_{3},\tau_{2},\tau_{1})G^{e(f)}
(\tau_{4},\tau_{3},\tau_{2},\tau_{1}),\nonumber\\
\end{eqnarray}
where the system response functions read \cite{Sun}
\begin{eqnarray}
G^{e}(\tau_{4},\tau_{3},\tau_{2},\tau_{1})=
\langle\Psi_{0}| \hat{\mu}_- e^{-iH_{\rm S}(\tau_{4}-\tau_{3})}\hat{\mu}_+\nonumber\\
\times e^{-iH_{\rm ph}(\tau_{3}-\tau_{2})}
\hat{\mu}_- e^{-iH_{\rm S}(\tau_{2}-\tau_{1})}\hat{\mu}_+
|\Psi_{0}\rangle,\label{G1s}
\end{eqnarray}
\begin{eqnarray}
G^{f}(\tau_{4},\tau_{3},\tau_{2},\tau_{1})=
\langle\Psi_{0}| \hat{\mu}_-  e^{-iH_{\rm S}(\tau_{4}-\tau_{3})}
\hat{\mu}_- \nonumber\\
\times e^{-iH_{\rm S}(\tau_{3}-\tau_{2})}
\hat{\mu}_+ e^{-iH_{\rm S}(\tau_{2}-\tau_{1})}\hat{\mu}_+
|\Psi_{0}\rangle.\label{G1d}
\end{eqnarray}
and by using the cumulant expansion, the bath response functions are written as\cite{Sun}
\begin{eqnarray}\label{Fi1}
F^{e}(\tau_{4},\tau_{3},\tau_{2},\tau_{1})&=&\exp \{ - [g(\tau_{2}-\tau_{1})-g(\tau_{3}-\tau_{1})\nonumber\\
&&+g(\tau_{4}-\tau_{1})+g(\tau_{3}-\tau_{2})\nonumber\\
&&-g(\tau_{4}-\tau_{2})+g(\tau_{4}-\tau_{3})]\},\nonumber\\
F^{f}(\tau_{4},\tau_{3},\tau_{2},\tau_{1})&=&\exp \{ - [-g(\tau_{2}-\tau_{1})+g(\tau_{3}-\tau_{1})\nonumber\\
&&+g(\tau_{4}-\tau_{1})+g(\tau_{3}-\tau_{2})\nonumber\\
&&+g(\tau_{4}-\tau_{2})-g(\tau_{4}-\tau_{3})]\}.
\end{eqnarray}
Herein, the lineshape functions are expressed taking the bath spectral density into account, that is \cite{Abramavicius,Sun},
\begin{eqnarray} \label{gt}
g(t)&=&\int_0^\infty d\omega\frac{D(\omega)}{\omega^2}[\coth{\frac{\beta\hbar\omega}{2}}(1-\cos{\omega t})\nonumber\\
&&+i(\sin{\omega t}-\omega t)],
\end{eqnarray}
where we adopt the Drude spectral density of the secondary bath,
\begin{equation}\label{spectrum0}
D(\omega)=2\eta \omega\frac{\gamma}{\omega^2+\gamma^2},
\end{equation}
and then derive the lineshape functions as \cite{Sun,Abramavicius,Mukamel}
\begin{eqnarray}\label{spectrum}
g(t)&=&\frac{\eta}{\gamma}\cot{\frac{\gamma\beta}{2}}[e^{-\gamma t}+\gamma t-1]-\nonumber\\
&&i\frac{\eta}{\gamma}[e^{-\gamma t}+\gamma t-1]+\frac{4\eta\gamma}{\beta}\sum_{n=1}^{\infty}\frac{e^{-\nu_n t}+\nu_n t-1}{\nu_n(\nu_n^2-\gamma^2)},\nonumber\\
\end{eqnarray}
with $\nu_n=2\pi n/\beta$ being the Matsubara frequencies. In this work, these parameters are taken as $\eta=0.15\omega_0$, $\gamma=0.03\omega_0$, and $k_{\rm B}T_{\rm eq}=0.15\omega_0$, with $\omega_0=$1500cm$^{-1}$.

Combining Eqs. (\ref{G1s})-(\ref{Fi1}) we get the excited-state response functions as \cite{Huynh,Sun}
\begin{eqnarray}\label{response functions}
&&R_1(\tau,T_{\rm w},t)=\sum_{ii_1i_2i_3}(\textbf{e}_4^*\cdot\vec{\mu}_{i_2}^*)(\textbf{e}_1\cdot\vec{\mu}_{i_3})(\textbf{e}_2^*\cdot\vec{\mu}^*_{i})\nonumber\\
&&(\textbf{e}_3\cdot\vec{\mu}_{i_1})A^*_{i_1i}(T_{\rm w})A_{i_2i_3}(\tau+T_{\rm w}+t)\nonumber\\
&&e^{-\frac 1 2\sum_q(|\lambda_{i_1q}(T_{\rm w})|^2+|\lambda_{i_2q}(\tau+T_{\rm w}+t)|^2)}\nonumber\\
&&e^{\lambda^*_{i_1q}(T_{\rm w})\lambda_{i_2q}(\tau+T_{\rm w}+t)e^{i\omega_qt}}F_1^e(\tau,T_{\rm w},t),\nonumber\\
&&R_2(\tau,T_{\rm w},t)=\sum_{ii_1i_2i_3}(\textbf{e}_4^*\cdot\vec{\mu}_{i_2}^*)(\textbf{e}_1^*\cdot\vec{\mu}_{i}^*)(\textbf{e}_2\cdot\vec{\mu}_{i_3})\nonumber\\
&&(\textbf{e}_3\cdot\vec{\mu}_{i_1})A^*_{i_1i}(\tau+T_{\rm w})A_{i_2i_3}(T_{\rm w}+t)\nonumber\\
&&e^{-\frac 1 2\sum_q(|\lambda_{i_1q}(\tau+T_{\rm w})|^2+|\lambda_{i_2q}(T_{\rm w}+t)|^2)}\nonumber\\
&&e^{\lambda^*_{i_1q}(\tau+T_{\rm w})\lambda_{i_2q}(T_{\rm w}+t)e^{i\omega_qt}}F_2^e(\tau,T_{\rm w},t),\nonumber\\
&&R_3(\tau,T_{\rm w},t)=\sum_{ii_1i_2i_3}(\textbf{e}_4^*\cdot\vec{\mu}_{i_2}^*)(\textbf{e}_1^*\cdot\vec{\mu}_{i}^*)(\textbf{e}_2\cdot\vec{\mu}_{i_1})\nonumber\\
&&(\textbf{e}_3\cdot\vec{\mu}_{i_3})A^*_{i_1i}(\tau)A_{i_2i_3}(t)\nonumber\\
&&e^{-\frac 1 2\sum_q(|\lambda_{i_1q}(\tau)|^2+|\lambda_{i_2q}(t)|^2)}\nonumber\\
&&e^{\lambda^*_{i_1q}(\tau)\lambda_{i_2q}(t)e^{i\omega_q(T_{\rm w}+t)}}F_3^e(\tau,T_{\rm w},t),\nonumber\\
&&R_4(\tau,T_{\rm w},t)=\sum_{ii_1i_2i_3}(\textbf{e}_4^*\cdot\vec{\mu}_{i}^*)(\textbf{e}_1\cdot\vec{\mu}_{i_3})(\textbf{e}_2^*\cdot\vec{\mu}^*_{i_2})\nonumber\\
&&(\textbf{e}_3\cdot\vec{\mu}_{i_1})A^*_{i_1i}(-t)A_{i_2i_3}(\tau)\nonumber\\
&&e^{-\frac 1 2\sum_q(|\lambda_{i_1q}(-t)|^2+|\lambda_{i_2q}(\tau)|^2)}\nonumber\\
&&e^{\lambda^*_{i_1q}(-t)\lambda_{i_2q}(\tau)e^{-i\omega_qT_{\rm w}}}F_4^e(\tau,T_{\rm w},t),\nonumber\\
\end{eqnarray}
where $\textbf{e}_4$ denotes the polarization of the local oscillator field,
$A_{i_1i}(t)$ represent the probability amplitude at time $t$ for the exciton at the state $|i_1\rangle$ with the initial state $|i\rangle$, and $\lambda_{i_1q}(t)$ are the corresponding displacement of phonon. $F_{1-4}^e(\tau,T_{\rm w},t)$ are the lineshape factors of the response functions $R_{1-4}$.
Assuming that the system-bath coupling is the same
for all excited states, we obtain the lineshape factors
within the second-order cumulant expansion, which are determined by \cite{Cho11,Abramavicius,Mukamel}
\begin{eqnarray}\label{lineshape}
F_1^{e}(\tau,T_{\rm w},t)&=&\exp[-g^*(t)-g(\tau)-g^*(T_{\rm w})+g^*(T_{\rm w}+t)\nonumber\\
&&+g(\tau+T_{\rm w})-g(\tau+T_{\rm w}+t)],\nonumber\\
F_2^{e}(\tau,T_{\rm w},t)&=&\exp[-g^*(t)-g^*(\tau)+g(T_{\rm w})-g(T_{\rm w}+t)\nonumber\\
&&-g^*(\tau+T_{\rm w})+g^*(\tau+T_{\rm w}+t)],\nonumber\\
F_3^{e}(\tau,T_{\rm w},t)&=&\exp[-g(t)-g^*(\tau)+g^*(T_{\rm w})-g^*(T_{\rm w}+t)\nonumber\\
&&-g^*(\tau+T_{\rm w})+g^*(\tau+T_{\rm w}+t)],\nonumber\\
F_4^{e}(\tau,T_{\rm w},t)&=&\exp[-g(t)-g(\tau)-g(T_{\rm w})+g(T_{\rm w}+t)\nonumber\\
&&+g(\tau+T_{\rm w})-g(\tau+T_{\rm w}+t)].
\end{eqnarray}
The higher excited-state response functions are given by\cite{Sun}
\begin{eqnarray}\label{response functions1}
&&R_1^*(\tau,T_{\rm w},t)=\sum_{\mbox{\tiny$\begin{array}{c}
ii_1i_2\\
i_3f\end{array}$}}(\textbf{e}_4^*\cdot\vec{\mu}_{i_1f}^*)(\textbf{e}_1^*\cdot\vec{\mu}_{i}^*)(\textbf{e}_2\cdot\vec{\mu}_{i_3})(\textbf{e}_3\cdot\vec{\mu}_{i_2f})\nonumber\\
&&A^*_{f(i_1i)}(0)A_{f(i_2i_3)}(t)e^{\sum_q\lambda^*_{f(i_1i),q}(0)\lambda_{f(i_2i_3),q}(t)}\nonumber\\
&&e^{-\frac 1 2\sum_q(|\lambda^*_{f(i_1i),q}(0)|^2+|\lambda_{f(i_2i_3),q}(t)|^2)}F_1^{f}(\tau,T_{\rm w},t),\nonumber\\
&&R_2^*(\tau,T_{\rm w},t)=\sum_{\mbox{\tiny$\begin{array}{c}
ii_1i_2\\
i_3f\end{array}$}}(\textbf{e}_4^*\cdot\vec{\mu}_{i_1f}^*)(\textbf{e}_1\cdot\vec{\mu}_{i_3})(\textbf{e}_2^*\cdot\vec{\mu}_{i}^*)(\textbf{e}_3\cdot\vec{\mu}_{i_2f})\nonumber\\
&&A^{\prime*}_{f(i_1i)}(0)A^{\prime}_{f(i_2i_3)}(t)e^{\sum_q\lambda^{\prime*}_{f(i_1i),q}(0)\lambda^{\prime}_{f(i_2i_3),q}(t)}\nonumber\\
&&e^{-\frac 1 2\sum_q(|\lambda^{\prime*}_{f(i_1i),q}(0)|^2+|\lambda^{\prime}_{f(i_2i_3),q}(t)|^2)}F_2^{f}(\tau,T_{\rm w},t).
\end{eqnarray}
The initial amplitudes of the higher excited-state are $A^*_{f(i_1i)}(0)=A^*_{i_1i}(\tau+T_{\rm w}+t)$, $A_{f(i_2i_3)}(0)=A_{i_2i_3}(T_{\rm w})$,
$A^{\prime*}_{f(i_1i)}(0)=A^{*}_{i_1i}(t+T_{\rm w})$, and $A^{\prime}_{f(i_2i_3)}(0)=A_{(i_2i_3)}(\tau+T_{\rm w})$, and the corresponding phonon displacements are $\lambda^*_{f(i_1i),q}(0)=\lambda^*_{i_1q}(\tau+T_{\rm w}+t)$, $\lambda_{f(i_2i_3),q}(0)=\lambda_{i_2q}(T_{\rm w})$, $\lambda^{\prime*}_{f(i_1i),q}(0)=\lambda^*_{i_1q}(t+T_{\rm w})$, and $\lambda^{\prime}_{f(i_2i_3),q}(0)=\lambda_{i_2q}(\tau+T_{\rm w})$. The lineshapes for $R_1^*$ and $R_2^*$ are correspondingly written as\cite{Cho11,Abramavicius,Sun,Meier}
\begin{eqnarray}\label{lineshape}
F_1^{f}(\tau,T_{\rm w},t)&=&\exp[g^*(t+T_{\rm w}+\tau)-g^*(T_{\rm w}+\tau)\nonumber\\
&&-g(t+T_{\rm w})-g^*(\tau)-g(t)+g(T_{\rm w})]\nonumber\\
F_2^{f}(\tau,T_{\rm w},t)&=&\exp[g^*(t+T_{\rm w})-g(t+T_{\rm w}+\tau)\nonumber\\
&&+g(T_{\rm w}+\tau)-g^*(T_{\rm w})-g(\tau)-g(t)].\nonumber\\
\end{eqnarray}

\section{2D spectrum}
For the third-order polarization $P^{(3)}(t)$, the outgoing-field directions with phase-matching condition ${\mathbf k}_I=-{\mathbf k}_1+{\mathbf k}_2+{\mathbf k}_3$ and ${\mathbf k}_{II}={\mathbf k}_1-{\mathbf k}_2+{\mathbf k}_3$ yield two contributions named rephasing (subscript R) and non-rephasing (subscript NR), respectively. In the impulsive limit, the two contributions of the third-order polarization read\cite{Huynh,Sun,Mukamel}
\begin{eqnarray}\label{rephasing}
P^{(3)}_{\rm R}(\tau,T_{\rm w},t)&\sim& -i[R_2(\tau,T_{\rm w},t)+R_3(\tau,T_{\rm w},t)\nonumber\\
&&-R_1^*(\tau,T_{\rm w},t)]
\end{eqnarray}
and
\begin{eqnarray}\label{non-rephasing}
P^{(3)}_{\rm NR}(\tau,T_{\rm w},t)&\sim& -i[R_1(\tau,T_{\rm w},t)+R_4(\tau,T_{\rm w},t)\nonumber\\
&&-R_2^*(\tau,T_{\rm w},t)].
\end{eqnarray}
The rephasing and non-rephasing 2D
PE spectra are subsequently evaluated by two-dimensional Fourier-Laplace transforms as \cite{Huynh,Sun,Mukamel}
\begin{eqnarray}\label{SR}
&&S_{\rm R}(\omega_\tau,T_{\rm w},\omega_t)  \nonumber\\
&&={\mathrm{Re}} \int_0^\infty \int_0^\infty dtd\tau\ iP_{\rm R}^{(3)}(\tau,T_{\rm w},t)
e^{-i\omega_\tau\tau+i\omega_tt},
\end{eqnarray}
\begin{eqnarray}\label{SNR}
&&S_{\rm NR}(\omega_\tau,T_{\rm w},\omega_t)  \nonumber\\
&&={\mathrm{Re}} \int_0^\infty \int_0^\infty dtd\tau\ iP_{\rm NR}^{(3)}(\tau,T_{\rm w},t)
e^{i\omega_\tau\tau+i\omega_tt}.
\end{eqnarray}
The total 2D spectrum is then defined by the sum of the two,
\begin{eqnarray}\label{St}
S(\omega_\tau,T_{\rm w},\omega_t)=S_{\rm R}(\omega_\tau,T_{\rm w},\omega_t)+S_{\rm NR}(\omega_\tau,T_{\rm w},\omega_t).
\end{eqnarray}


\begin{thebibliography}{Markovian adiabaticity appendix}
\bibitem{note1} The corresponding e$_1^{\prime}$ excitation parts are not signed due to the overlap of the peaks.
\end{thebibliography}
\end{document}